# Severity Assessment of Coronavirus Disease 2019 (COVID-19) Using Quantitative Features from Chest CT Images

Zhenyu Tang[1,#], Wei Zhao[2,#], Xingzhi Xie[2], Zheng Zhong[3,4], Feng Shi[5], Jun Liu[2,6,*], Dinggang Shen[5,*]

**Affiliations:**

[1] Beijing Advanced Innovation Center for Big Data and Brain Computing, Beihang University, Beijing 100191, China

[2] Department of Radiology, The Second Xiangya Hospital, Central South University, Changsha, Hunan Province 410011, China

[3] Department of Radiology, First hospital of Changsha, Hunan Province, 410005, China

[4] Changsha public health treatment center, Hunan Province, 410153, China

[5] Shanghai United Imaging Intelligence Co., Ltd., Shanghai 200232, China

[6] Department of Radiology Quality Control Center, Changsha, Hunan Province, 410008, China

[#] Z. Tang and W. Zhao contributed equally as first authors of this paper

[*] Corresponding authors:

Jun Liu, Department of Radiology, Second Xiangya Hospital, Central South University, Changsha, Hunan Province 410011, China and Department of Radiology Quality Control Center, Changsha, Hunan Province, 410008, China. junliu123@csu.edu.cn.

Dinggang Shen, Shanghai United Imaging Intelligence Co., Ltd., Shanghai 200232, China, dinggang.shen@gmail.com.

## Abstract

**Background**: Chest computed tomography (CT) is recognized as an important tool for COVID-19 severity assessment. As the number of affected patients increase rapidly, manual severity assessment becomes a labor-intensive task, and may lead to delayed treatment.

**Purpose**: Using machine learning method to realize automatic severity assessment (non-severe or severe) of COVID-19 based on chest CT images, and to explore the severity-related features from the resulting assessment model.

**Materials and Method**: Chest CT images of 176 patients (age 45.3$\pm$16.5 years, 96 male and 80 female) with confirmed COVID-19 are used, from which 63 quantitative features, e.g., the infection volume/ratio of the whole lung and the volume of ground-glass opacity (GGO) regions, are calculated. A random forest (RF) model is trained to assess the severity (non-severe or severe) based on quantitative features. Importance of each quantitative feature, which reflects the correlation to the severity of COVID-19, is calculated from the RF model.

**Results**: Using three-fold cross validation, the RF model shows promising results, i.e., 0.933 of true positive rate, 0.745 of true negative rate, 0.875 of accuracy, and 0.91 of area under receiver operating characteristic curve (AUC). The resulting importance of quantitative features shows that the volume and its ratio (*with respect to* the whole lung volume) of ground glass opacity (GGO) regions are highly related to the severity of COVID-19, and the quantitative features calculated from the right lung are more related to the severity assessment than those of the left lung.

**Conclusion**: The RF based model can achieve automatic severity assessment (non-severe or severe) of COVID-19 infection, and the performance is promising. Several quantitative features, which have the potential to reflect the severity of COVID-19, were revealed.

## Introduction

Since the end of year 2019, a novel coronavirus disease 2019 (COVID-19) was detected and has rapidly spread out worldwide. Recently, the world health organization (WHO) has declared that COVID-19 becomes a global pandemic [1]. Over thirty thousands of people around the world have been confirmed as COVID-19 infection, and more people are suspected. Reverse transcription polymerase chain reaction (RT-PCR) is one of the standard diagnostic methods, but it faces the inherent disadvantage of false negative positive due to many factors, e.g., quality, stability, reproducibility and possible insufficient viral material in specimens [2]. Chest computed tomography (CT) has been identified as an important complementary tool for the diagnosis of COVID-19. Comparing to RT-PCR test, chest CT is relatively easy to operate and can provide fast diagnosis; moreover, chest CT has high sensitivity for screening COVID-19 infection [3]. Therefore, chest CT has become an indispensable tool in the screening and severity assessment of COVID-19 in China. However, since the explosively increased confirmed and suspected cases of COVID-19, manually assessing the severity of COVID-19 based on chest CT images becomes a labor-intensive task, potentially leading to the delay of patient isolation and treatment.

Although automatic severity assessment for COVID-19 using chest CT is highly desired, very few studies focusing on this topic has been reported. Most of current studies on chest CT images are about the findings of image characteristics related to COVID-19, e.g., the ground glass opacity (GGO) and consolidation regions in chest CT images [4, 5]. However, to the best of our knowledge, no study has worked on automatic severity assessment of COVID-19 using chest CT images.

In this paper, we build a random forest (RF) based model for COVID-19 severity assessment. The input of the model is the quantitative features calculated from chest CT images, and the output is the severity of COVID-19 (non-severe or severe). In order to reduce the redundant and unrelated quantitative features, in the training stage, the importance of all quantitative features in the context of COVID-19 severity assessment is calculated, and a subset of quantitative features that are of great importance is selected to train the RF model. The performance of the resulting RF model is found to be very promising.

## Material and methods

This multi-center study was approved by Medical Ethical Committee (approved number: 2020002), which waived the requirement for patients' informed consent referring to the Council for International Organizations of Medical Sciences (CIOMS) guideline.

*Chest CT images*

In total, chest CT images of 176 patients (age 45.3±16.5, 96 male and 80 female) with confirmed COVID-19 (RT-PCR test positive) are included in this study. All chest CT images were captured in seven hospitals with different types of CT scanners, including Philips (Ingenuity CT iDOSE4), Siemens (Somatom perspective), GE (Bright speed S), Hitachi (ECLOS) and Anke (ANATOM 16HD). The scanning parameters are as follows: 120 kVp, 100-200 mAs, pitch 0.75-1.5 and collimation 1-5 mm. All imaging data were reconstructed by using a medium sharp reconstruction algorithm with a thickness of 1-5 mm.

*The clinical condition of COVID-19*

For each patient, the severity of COVID-19 is determined following the guideline of 2019-nCoV (trial version 7) issued by the China National Health Commission, and the severity includes four types: mild, common, severe, and critical [6]. However, the number of patients with mild and critical types are small as compared to those with common and severe types. In order to avoid the imbalanced issue, we categorize patients into two groups: non-severe group (mild and common) and severe group (severe and critical), and then do two-type severity assessment (non-severe or severe), i.e., binary classification. It is worth noting that the binary classification is reasonable in clinical practice, as patients with non-severe and severe types are different in treatment regimen. Of these 176 patients, 121 are identified as non-severe, and 55 are identified as severe. Fig. 1 shows examples of the chest CT images from 2 non-severe patients and 2 severe patients.

*Quantitative features in chest CT images*

In this study, a COVID-19 chest CT analysis tool (uAI-Discover-NCP) [7], which is developed by Shanghai United Imaging Intelligence Co., Ltd., is used to calculate the quantitative features from the chest CT images. This tool is based on deep learning, where a VB-net [7] is adopted to fulfill accurate segmentation of lung as well as infection regions from chest CT images. It is evaluated on 300 subjects with 0.916 Dice similarity with manual delineations. According to segmentation results, quantitative features, which are potentially related to COVID-19, are calculated. Specifically, the lung is segmented and divided into five lung lobes, i.e., superior/middle/inferior lobes of the right lung ($\mathrm{RB}_{\mathrm{S|M|I}}$) and superior/inferior lobes of the left lung ($\mathrm{LB}_{\mathrm{S|I}}$), and 18 lung segments, where 10 segments in the right lung ($\mathrm{RS}_{1-10}$) and 8 segments in the left lung ($\mathrm{LS}_{1-8}$). The infection volume ($\boldsymbol{IV}$) and ratio ($\boldsymbol{IR}$) of the whole lung (WL), right/left lung (RL/LL), and each lobe/segment are calculated as quantitative features, as defined below:

$$\boldsymbol{IV}(x) = \boldsymbol{V}(x_{\mathrm{infect}}), \quad \boldsymbol{IR}(x) = \frac{\boldsymbol{IV}(x)}{\boldsymbol{V}(x)} = \frac{\boldsymbol{V}(x_{\mathrm{infect}})}{\boldsymbol{V}(x)} \qquad (1)$$

where $\boldsymbol{V}(.)$ is the volume of input region, $x \in \{\mathrm{WL}, \mathrm{RL}, \mathrm{LL}, \mathrm{RB}_{\mathrm{S|M|I}}, \mathrm{LB}_{\mathrm{S|I}}, \mathrm{RS}_{1-10}, \mathrm{LS}_{1-8}\}$ and $x_{\mathrm{infect}}$ is the infected regions in $x$. Moreover, regions within the HU ranges of $-\infty$ to -750 ($\mathrm{HU}_{[-\infty,-750]}$), -750 to -300 ($\mathrm{HU}_{[-750,-300]}$), -300 to 50 ($\mathrm{HU}_{[-300,50]}$), and 50 to $+\infty$ ($\mathrm{HU}_{[50,+\infty]}$) are also considered in our study. These four HU ranges correspond to normal lung regions, ground glass opacity (GGO) regions, consolidation regions, and regions of vessel calcification, respectively. Volumes and ratios of these four regions are also calculated and adopted as quantitative features. It is worth noting that, the ratio of the region $x$ within each HU range is calculated by $\boldsymbol{R}(x) = \frac{\boldsymbol{V}(x)}{\boldsymbol{V}(\mathrm{WL})}$ (i.e., *with respect to* the volume of the whole lung). These 60 quantitative features together with the volumes of the whole/right/left lung form the final quantitative feature set (i.e., 63 quantitative features in total). A summary of these 63 quantitative features is presented in Table 1.

*Random forest based severity assessment*

Random forest (RF) is a kind of ensemble learning method [8], where training samples for building each decision tree [9] are randomly selected (bootstrap samples) [10], and the final result is obtained by majority voting or averaging the prediction results from multiple decision trees. Since RF is robust to noise [11] and relatively insensitive to small amount of training samples [12], it has been widely used for classification and regression tasks in medical field [13-15]. In this study, the classification and regression tree (CART) is adopted to build the RF model. The Gini index is used in CART to measure the split quality at each node, which is defined as:

$$\boldsymbol{G}(x) = \sum_{c=1}^{C} \boldsymbol{P}_c(x) \sum_{c' \neq c} \boldsymbol{P}_{c'}(x) = 1 - \sum_{c=1}^{C} \boldsymbol{P}_c^2(x) \quad (2)$$

where $x$ is the sample set at a node, $\boldsymbol{P}_c(x)$ is the probability of samples in $x$ belonging to class $c$, $C$ is the number of all classes, and $C = 2$ for binary classification. At each node, the CART chooses a feature, which has the smallest Gini index of $\frac{|x_{\mathrm{left}}|}{|x|}\boldsymbol{G}(|x_{\mathrm{left}}) + \frac{|x_{\mathrm{right}}|}{|x|}\boldsymbol{G}(|x_{\mathrm{right}})$ of all features ($|.|$ is the sample size), to split the samples in $x$ into two subgroups (i.e., the left and right child nodes containing samples $x_{\mathrm{left}}$ and $x_{\mathrm{right}}$, respectively).

Considering the relatively imbalanced samples, i.e., the ratio between non-severe and severe samples is 11:5, a weighted random forest (WRF) strategy [16] is applied. Specifically, instead of assigning equal weights to training samples in traditional RF (i.e., setting sample weight as 1), the sample weights in WRF are inversely proportional to the sample numbers of their classes, i.e., assigning large weights to samples in minority classes. Fig. 2 provides the whole workflow of the RF based severity assessment for COVID-19.

## Results

Three-fold cross-validation is performed to evaluate the RF based severity assessment model. In each case, the training samples are divided into training (70%) and validation (30%) samples. It is worth noting that, the ratio between non-severe and severe samples in the training, validation, and testing samples is relatively consistent with that in all 176 samples (i.e., around 11:5).

The RF model contains 500 decision trees. The input of the RF model includes 63 quantitative features calculated from chest CT images, and the output is the severity of COVID-19 (i.e., non-severe or severe). To explore the relations of 63 quantitative features to the severity of COVID-19, importance of each quantitative feature is also calculated based on the RF model. The importance of each quantitative feature ($F_i, i = 1, \ldots, 63$) is related to the reduced Gini index in the RF model using $F_i$, which reflects the capacity of $F_i$ in discriminating non-severe and severe types. Specifically, the reduced Gini index ($\boldsymbol{G}_{\mathrm{RD}}$) of $F_i$ is calculated by:

$$\boldsymbol{G}_{\mathrm{RD}}(F_i) = \frac{1}{|\mathrm{node}^i|} \sum_{x^i \in \mathrm{node}^i} \boldsymbol{G}(x^i) - \boldsymbol{G}(x^i_{\mathrm{left}}) - \boldsymbol{G}(x^i_{\mathrm{right}}) \tag{3}$$

where $\mathrm{node}^i$ is the nodes in the RF models that use quantitative feature $F_i$ to split samples $x^i$ in $\mathrm{node}^i$ into $x^i_{\mathrm{left}}$ and $x^i_{\mathrm{right}}$. Based on $\boldsymbol{G}_{\mathrm{RD}}(F_i)$, the importance of $F_i$ can be defined as:

$$\boldsymbol{IP}(F_i) = \frac{\boldsymbol{G}_{\mathrm{RD}}(F_i)}{\sum_{i=1}^{63} \boldsymbol{G}_{\mathrm{RD}}(F_i)}. \tag{4}$$

According to the resulting importance of quantitative features, top $K$ quantitative features of the great importance are selected and used to build the RF model. To get the proper hyper-parameter $K$, a grid search strategy is adopted. Specifically, in the experiment, besides the RF model using 63 quantitative features (i.e., K63), five more RF models with $K = 50, 40, 30, 20, 10$ are built, respectively, and their performance is evaluated and compared using validation samples. Details of the evaluation results of these RF models and their corresponding receiver operating characteristic (ROC) curves, together with area under ROC curve (AUC) values, are presented in Fig. 3.

It could be observed that the K30 RF model (i.e., $K = 30$) has the best performance using the validation samples. Therefore, the K30 RF model is chosen as the final severity assessment model and applied to the testing samples. Specifically, the evaluation result of K30 RF model using the testing samples are 0.933 (TPR), 0.745 (TNR) and 0.875 (Accuracy). The ROC curve of K30 RF model evaluated with the testing samples is shown in Fig. 4, and the AUC of the ROC curve is 0.91.

Table 2 shows the top 30 quantitative features of the great importance to the severity. The volume and ratio of the region within the HU range of [-700,-300], corresponding to the region of ground glass opacity (GGO) in chest CT images, is at the top rank. An interesting finding is that quantitative features calculated from right lung are relatively more important than those of left lung (Table 2).

## Discussions

In this paper, a random forest (RF) based automatic severity assessment of COVID-19 model was presented; and, to the best of our knowledge, this is the first work using machine learning method to perform severity assessment of COVID-19. Sixty-three quantitative features, including infection volume/ratio of the whole lung as well as the volumes of ground-glass opacity (GGO) regions, were calculated from the chest CT images and used as input of the RF model. The output of the RF model is the corresponding severity (non-severe or severe). In addition, the importance of 63 quantitative features was also calculated from the RF model, and different number of quantitative features, which are of the top importance, were further selected and used to train the RF model. The RF model using 30 quantitative features achieved the best performance using the validation samples, and the evaluation results using the testing samples are promising, i.e., TPR, TNR and Accuracy are 0.933, 0.745, and 0.875, respectively, along with AUC of 0.91.

Among these 30 quantitative features, the volumes of ground glass opacity (GGO) regions (HU range of -700 to -300) and their ratios (*with respect to* the whole lung volume) are the most important features contributed to the severity of COVID-19. This is contrary to several previous studies, which reported that consolidation regions were more like to be seen in patients in ICU (Intensive Care Unit) or later stage of the disease [17]. However, in this study, we only included initial CT images, indicating that GGOs might be the predominant findings than consolidation regions at baseline [18, 19]. In our cohort, the volumes and ratios of GGO regions are indeed greater than those of consolidation regions with $p = 3.28 \times 10^{-7}$ using pairwise t-test. Fig. 5 shows the ratios of GGO regions, i.e., HU range of [-700, -300], and consolidation regions, i.e., HU range of [-300, 50], of 176 patients. Moreover, the infection volume and ratio of the whole lung are also highly related to the severity of COVID-19; this finding further proved the previous study [20]. Another interesting finding is that the quantitative features of the right lung lobes were more relevant to the severity of COVID-19 than those of the left lung lobes. This may be explained by specific anatomical structure of the trachea and bronchi (i.e., short and straight), and the virus might easily infect this location. Please note that this finding is consistent with previous studies related to COVID-19 pneumonia baseline [18] and H7N9 baseline [21].

The limitation of this study is that two types of the COVID-19 severity (non-severe and severe) were used (i.e., with binary classification), instead of four types (i.e., mild, common, severe and critical), because of limited numbers of patients with mild and critical types of COVID-19, respectively. In this study, mild and common were regarded as non-severe, while severe and critical were combined as severe.

In the future, we will extend the severity types by collecting chest CT images from more patients through multi-center collaborations.

**Table 1: Summary of 63 quantitative features calculated from chest CT images.**

| ID | Quantitative feature | Descriptions |
|---|---|---|
| 1-3 | $\boldsymbol{V}(\mathrm{WL})$, $\boldsymbol{V}(\mathrm{RL})$, $\boldsymbol{V}(\mathrm{LL})$ | Volumes of the whole/right/left lung |
| 4-5 | $\boldsymbol{IV}(\mathrm{WL})$, $\boldsymbol{IR}(\mathrm{WL})$ | Infection volume in the whole lung and its ratio *with respect to* the whole lung |
| 6-7 | $\boldsymbol{IV}(\mathrm{RL})$, $\boldsymbol{IR}(\mathrm{RL})$ | Infection volume in the right lung and its ratio *with respect to* the right lung |
| 8-9 | $\boldsymbol{IV}(\mathrm{LL})$, $\boldsymbol{IR}(\mathrm{LL})$ | infection volume in the left lung and its ratio *with respect to* the left lung |
| 10-15 | $\boldsymbol{IV}(\mathrm{RB}_{\mathrm{S\|M\|I}})$, $\boldsymbol{IR}(\mathrm{RB}_{\mathrm{S\|M\|I}})$ | Infection volume in the superior/middle/ inferior lobe of the right lung and its ratio *with respect to* the corresponding lobes of the right lung |
| 16-19 | $\boldsymbol{IV}(\mathrm{LB}_{\mathrm{S\|I}})$, $\boldsymbol{IR}(\mathrm{LB}_{\mathrm{S\|I}})$ | Infection volume in the superior/inferior lobe of the left lung and its ratio *with respect to* the corresponding lobes of the left lung |
| 20-39 | $\boldsymbol{IV}(\mathrm{RS}_{1-10})$, $\boldsymbol{IR}(\mathrm{RS}_{1-10})$ | Infection volumes in each segment (1-10) of the right lung and its ratio *with respect to* the corresponding segment of the right lung |
| 40-55 | $\boldsymbol{IV}(\mathrm{LS}_{1-8})$, $\boldsymbol{IR}(\mathrm{LS}_{1-8})$ | Infection volume in each segment (1-8) of the left lung and its ratio *with respect to* the corresponding segment of the left lung |
| 56-57 | $\boldsymbol{V}(\mathrm{HU}_{[-\infty,-750]})$, $\boldsymbol{R}(\mathrm{HU}_{[-\infty,-750]})$ | Volume of normal lung regions and its ratio *with respect to* the volume of the whole lung |
| 58-59 | $\boldsymbol{V}(\mathrm{HU}_{[-750,-300]})$, $\boldsymbol{R}(\mathrm{HU}_{[-750,-300]})$ | Volume of ground glass opacity (GGO) regions and its ratio *with respect to* the volume of the whole lung |
| 60-61 | $\boldsymbol{V}(\mathrm{HU}_{[-300,50]})$, $\boldsymbol{R}(\mathrm{HU}_{[-300,50]})$ | Volume of consolidation regions and its ratio *with respect to* the volume of the whole lung |
| 62-63 | $\boldsymbol{V}(\mathrm{HU}_{[50,+\infty]})$, $\boldsymbol{R}(\mathrm{HU}_{[50,+\infty]})$ | Volume of vessel calcification regions and its ratio *with respect to* the volume of the whole lung |

**Table 2: Rank of the top 30 quantitative features according to their importance in severity assessment of COVID-19.**

| Rank | Quantitative feature | Rank | Quantitative feature | Rank | Quantitative feature |
|---|---|---|---|---|---|
| 1 | $\boldsymbol{V}(\mathrm{HU}_{[-\infty,-750]})$ | 11 | $\boldsymbol{IR}(\mathrm{RS}_5)$ | 21 | $\boldsymbol{IV}(\mathrm{RS}_{10})$ |
| 2 | $\boldsymbol{R}(\mathrm{HU}_{[-750,-300]}$ | 12 | $\boldsymbol{IR}(\mathrm{RS}_{10})$ | 22 | $\boldsymbol{IV}(\mathrm{RB}_{\mathrm{S}})$ |
| 3 | $\boldsymbol{IV}(\mathrm{RL})$ | 13 | $\boldsymbol{IR}(\mathrm{RS}_1)$ | 23 | $\boldsymbol{V}(\mathrm{HU}_{[-\infty,-750]})$ |
| 4 | $\boldsymbol{IV}(\mathrm{WL})$ | 14 | $\boldsymbol{V}(\mathrm{HU}_{[50,+\infty]})$ | 24 | $\boldsymbol{IV}(\mathrm{RB}_{\mathrm{M}})$ |
| 5 | $\boldsymbol{IR}(\mathrm{WL})$ | 15 | $\boldsymbol{IR}(\mathrm{LS}_7)$ | 25 | $\boldsymbol{IR}(\mathrm{RS}_7)$ |
| 6 | $\boldsymbol{IV}(\mathrm{RB}_{\mathrm{I}})$ | 16 | $\boldsymbol{IR}(\mathrm{RS}_9)$ | 26 | $\boldsymbol{IV}(\mathrm{LL})$ |
| 7 | $\boldsymbol{IR}(\mathrm{RB}_{\mathrm{I}})$ | 17 | $\boldsymbol{IV}(\mathrm{RS}_5)$ | 27 | $\boldsymbol{IR}(\mathrm{LS}_8)$ |
| 8 | $\boldsymbol{IR}(\mathrm{RL})$ | 18 | $\boldsymbol{IV}(\mathrm{RS}_8)$ | 28 | $\boldsymbol{IV}(\mathrm{RS}_4)$ |
| 9 | $\boldsymbol{IV}(\mathrm{RS}_1)$ | 19 | $\boldsymbol{IR}(\mathrm{RS}_8)$ | 29 | $\boldsymbol{IV}(\mathrm{RS}_7)$ |
| 10 | $\boldsymbol{IV}(\mathrm{LS}_7)$ | 20 | $\boldsymbol{V}(\mathrm{HU}_{[-300,50]})$ | 30 | $\boldsymbol{IR}(\mathrm{RB}_{\mathrm{M}})$ |

$\mathrm{HU}_{[a,b]}$: regions within the HU range of $[a, b]$; WL: whole lung; RL: right lung; LL: left lung;
$\mathrm{RB}_{\mathrm{S|M|I}}$: sup./mid./inf. lobe of the right lung; $\mathrm{RS}_{1-10}$: segments 1-10 of the right lung; $\mathrm{LS}_{1-8}$: segments 1-8 of the left lung.

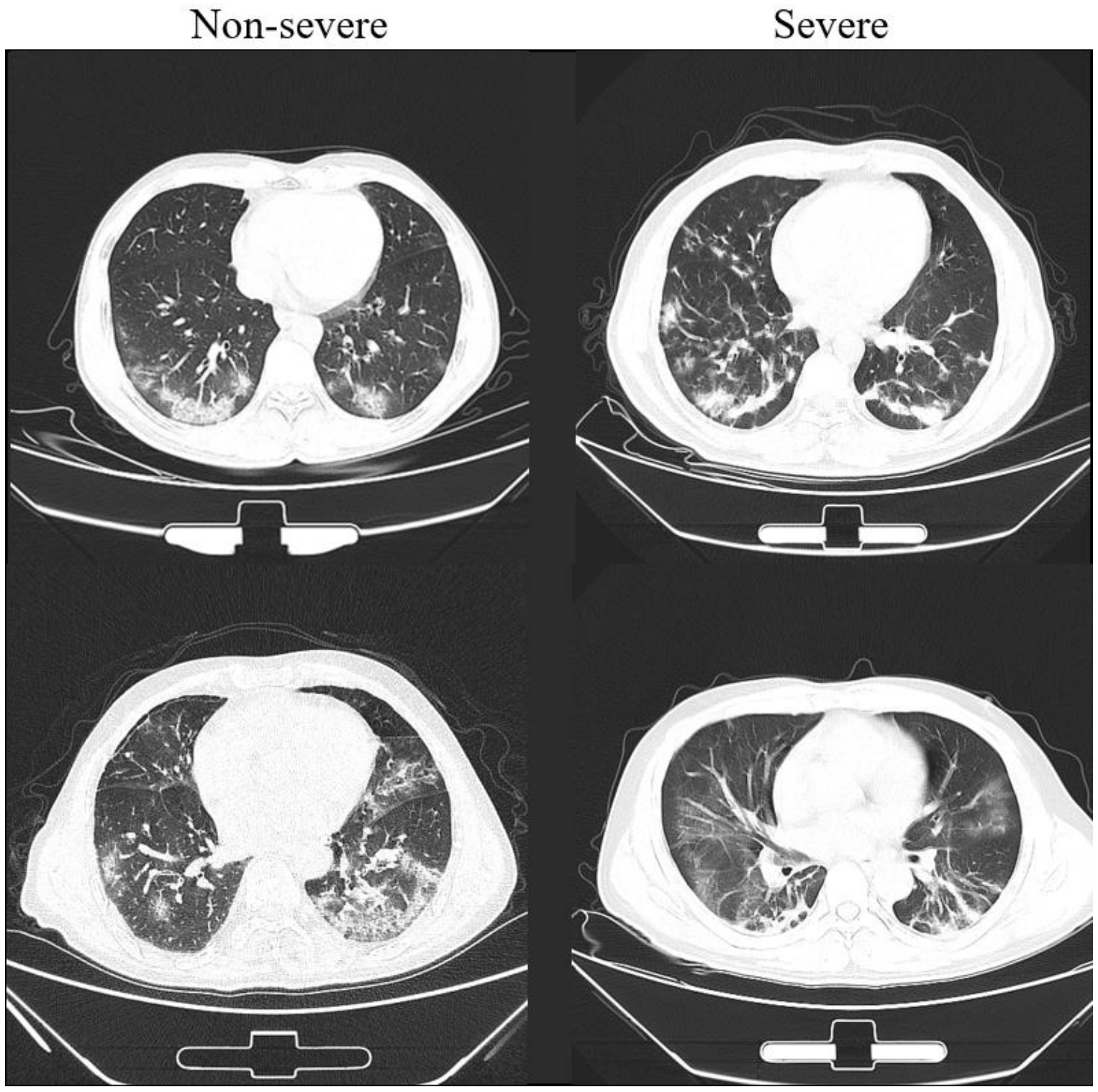

**Fig. 1. Chest CT examples of (left) 2 patients of non-severe infection COVID-19, and (right) 2 patients of severe COVID-19 infection.**

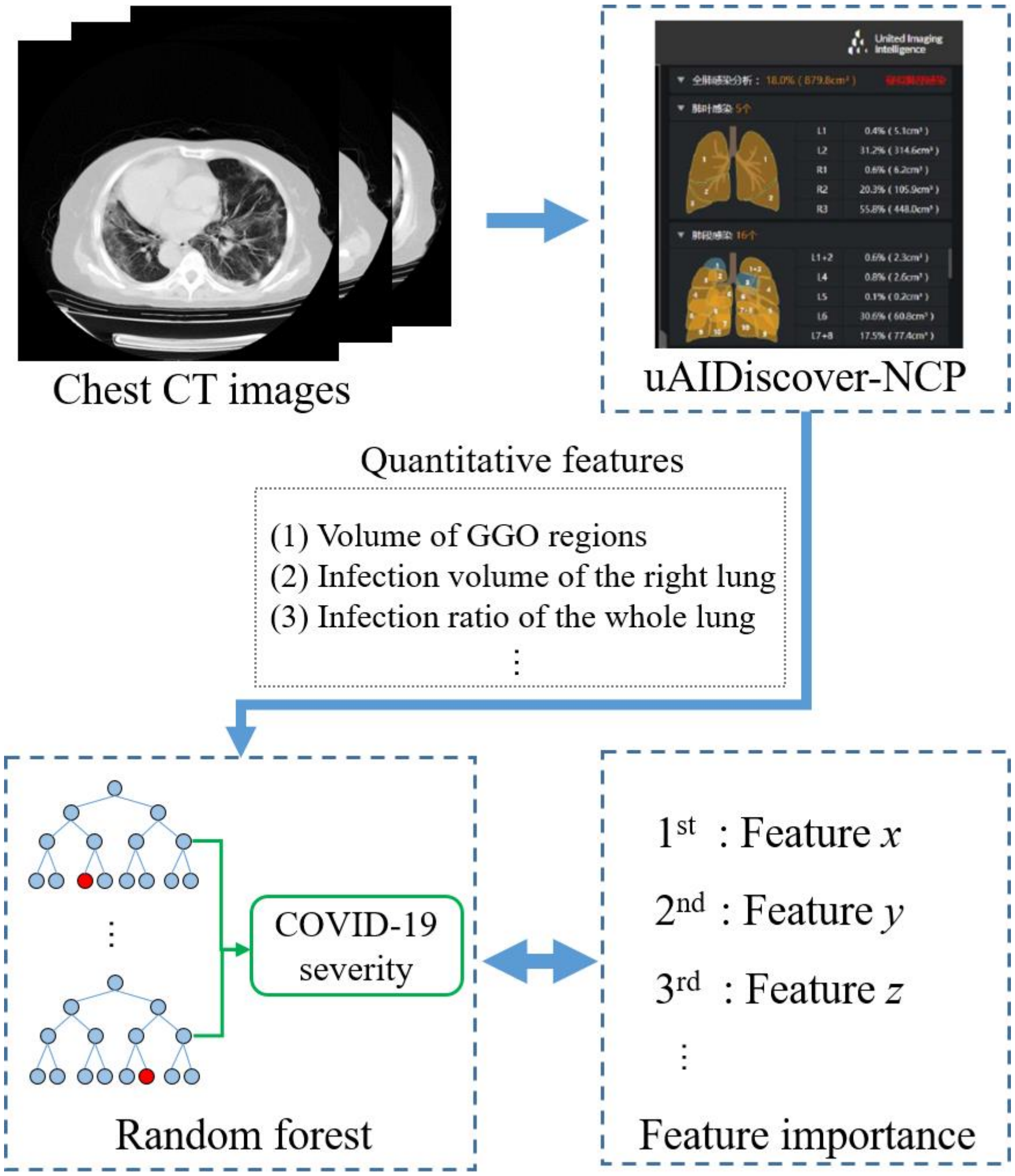

**Fig. 2. Workflow of the random forest based severity assessment for COVID-19.**

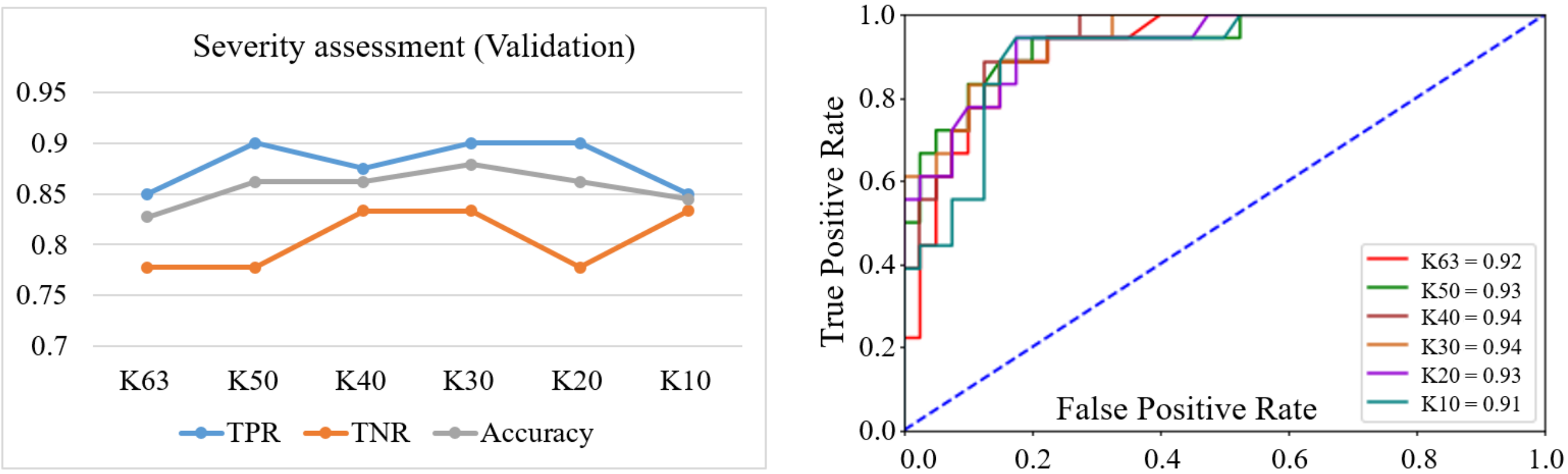

**Fig. 3. True positive rate (TPR), true negative rate (TNR), and Accuracy using RF models trained with different number of quantitative features (left), and the receiver operating characteristic (ROC) curves of these RF models (right). K63 stand for the RF models using 63 quantitative features, which is the same for other models.**

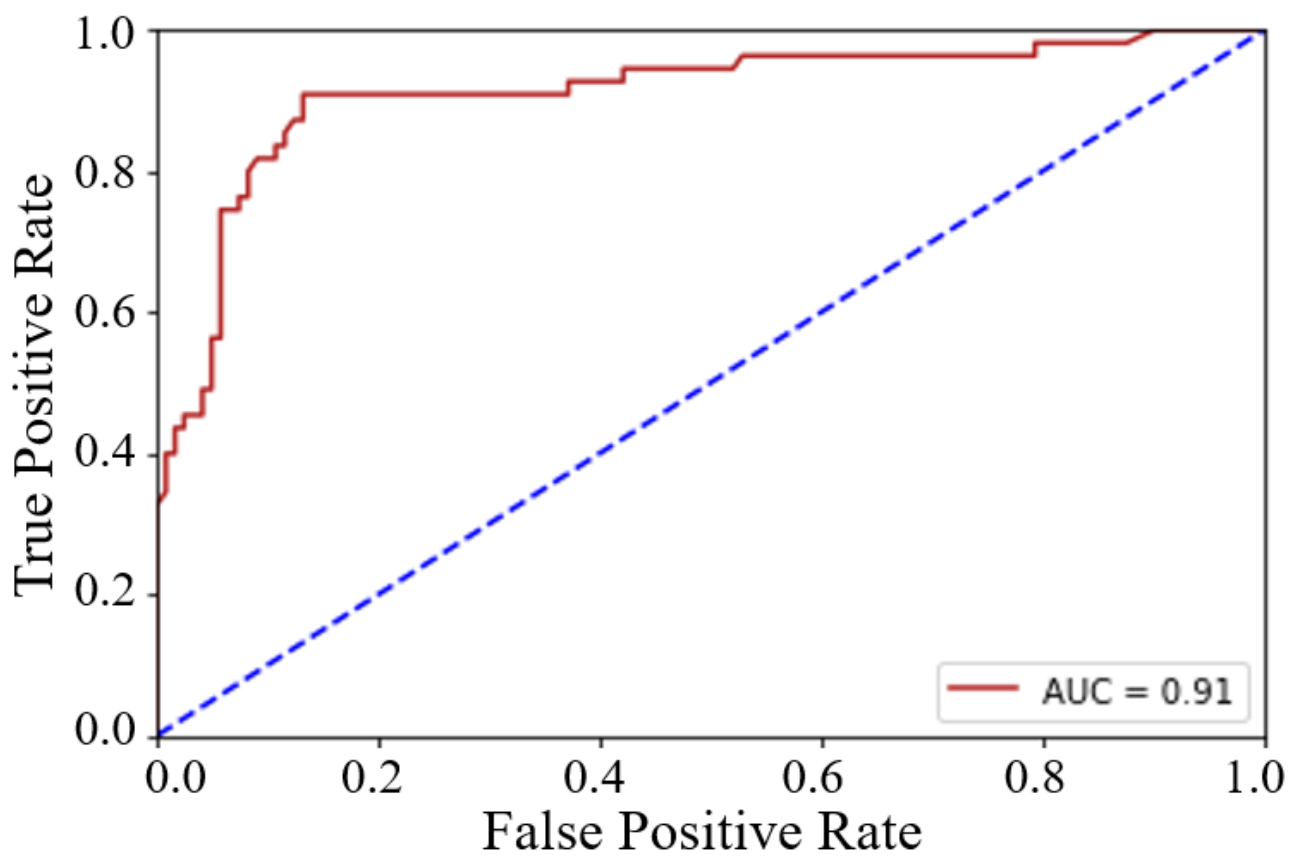

**Fig. 4. The receiver operating characteristic (ROC) curves of the K30 RF model (trained with top 30 important quantitative features) using the testing samples.**

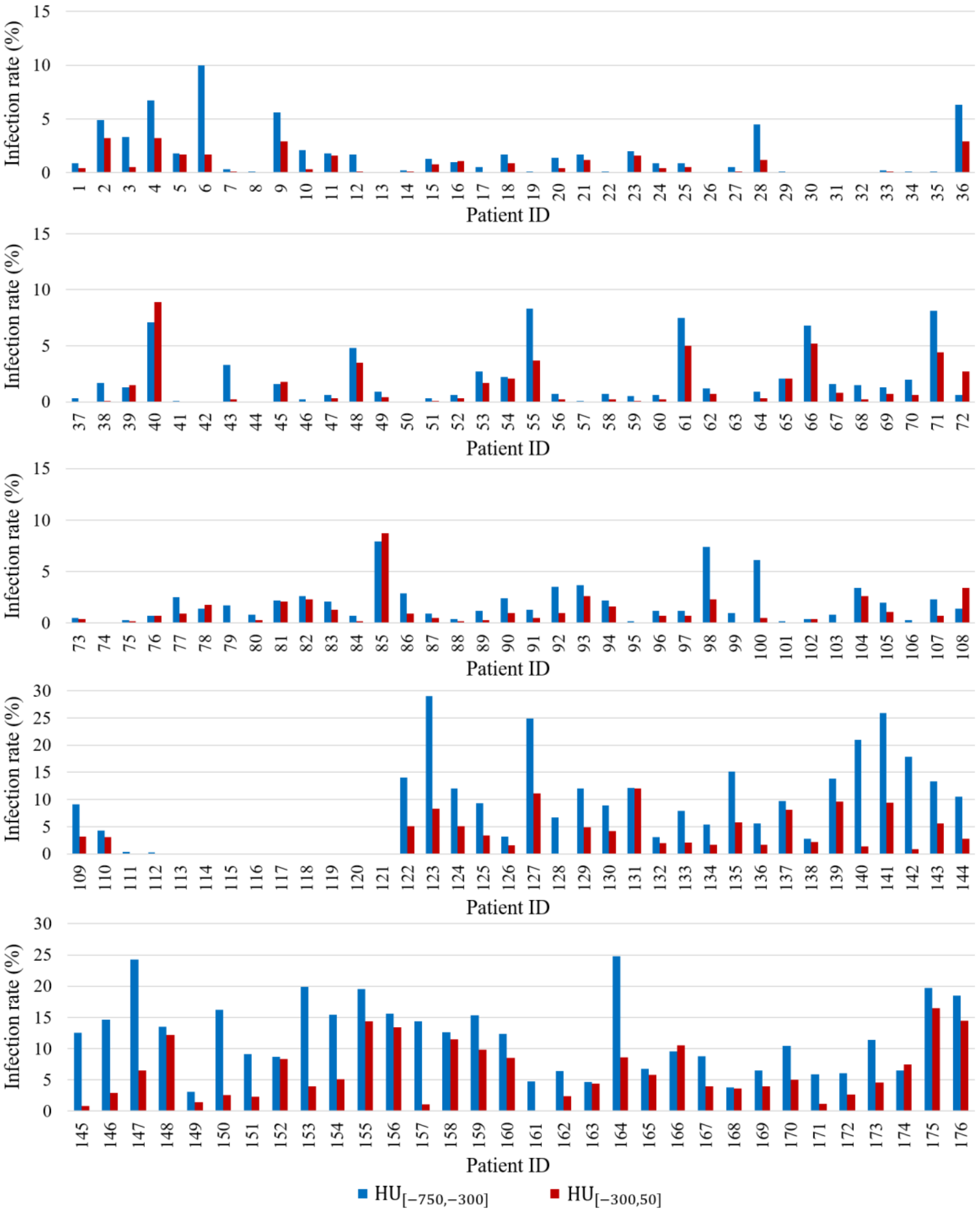

**Fig. 5. Ratios of GGO ($HU_{[-750,-300]}$) and consolidation ($HU_{[-300,50]}$) (*with respect to* the whole lung volume) in the chest CT images of 176 patients used in the study. Patient IDs 1-121 are non-severe, and Patient IDs 122-176 are severe.**